\documentclass[onecolumn,showpacs]{revtex4}
\usepackage{amssymb}
\usepackage{makeidx}
\usepackage{amsmath}
\usepackage{graphicx}
\usepackage{epstopdf}
\usepackage{dcolumn}
\usepackage{bm}

\input{tcilatex}
\begin{document}

\title{Multiphoton cross sections of conductive electrons stimulated
bremsstrahlung in doped bilayer graphene}
\author{A.G. Ghazaryan}
\email{amarkos@ysu.am}
\author{Kh.V. Sedrakian}
\affiliation{Centre of Strong Fields, Yerevan State University, 1 A. Manukian, Yerevan
0025, Armenia }
\date{\today }

\begin{abstract}
The quantum theory of multiphoton stimulated bremsstrahlung of charged
carriers on an arbitrary electrostatic potential of impurity ion in doped
bilayer graphene at the presence of coherent electromagnetic radiation is
developed. A terahertz wave field is considered exactly, while the
electrostatic potential of doped ions as a perturbation. The essentially
nonlinear response of bilayer graphene to a pump wave and significant
differences from the case of a single layer graphene are shown, which can be
associated to nonlinear parabolic dispersion. The latter opens new way to
manipulate with the electronic transport properties of conductive electrons
of bilayer graphene by coherent radiation field of terahertz or
near-infrared frequencies.
\end{abstract}

\maketitle

%%%%%%%%%%%%%%%%%%%%%%%%%%%%%%%%%%%%%%%%%%%%%%%%%%%%%%%%%%%%% 

%>>>> Include a list of up to six keywords after the abstract

%%%%%%%%%%%%%%%%%%%%%%%%%%%%%%%%%%%%%%%%%%%%%%%%%%%%%%%%%%%%%

\section{Introduction}

More than twenty years single layer graphene (SG) \cite{1,2} has attracted
the giant interest for its unusual electronic transport and the
characteristics of relativistic charge carriers behaving like massless
chiral Dirac fermions \cite{5,6,book,7a,7aa,7aaa,Mer2,Mer3,Mer4,Mer5,Mer6,Xu}. Bilayer graphene (BG) are
consisted of two graphene AB stacked single layers. If the SG graphene band
structure has a linear dispersion, BG has a quadratic dispersion \cite%
{2a,KoshinoAndo,2e,3b,1a} in the low-energy mode, which makes it similar to
2D semiconductor systems \cite{semicond}, except for the absence of an
energy gap. The BG is useful for both technological and fundamental
applications \cite{1a,2a,2e,KoshinoAndo,3b}, \cite%
{3a,3c,Screening,Doped}. In
particular, the BG is of current interest because of a tunable bandgap \cite%
{3c}, \cite{6c,7c,8c}, and extraordinary quantum Hall physics with an
eight-fold degenerate zero energy Landau level \cite{3b}, and anomalous
disorder and charge-carrier scattering \cite{Nilsson1,Nilsson2,BLG}.
Moreover, BG is of interest by considerably richer properties at multiphoton
resonant excitations and high harmonic generation \cite{Mer4}, \cite%
{Merarxiv1}, \cite{Merarxiv2}. However, little is known about the effective
absorption of coherent electromagnetic (EM) radiation via the multiphoton
stimulated bremsstrahlung (SB) process \cite{book} on charged impurity ions
in BG. Among the induced by external radiation field processes SB is one of
the first stimulated effects at laser-matter interaction, revealed after the
invention of lasers \cite{12a}, \cite{13a}. The SB process is the main
mechanism of energy exchange between charged particles and a plane
monochromatic wave in plasma-like medium for real absorption-emission
processes.

As it is known, the interaction of electron with the EM wave in initially
nonrelativistic plasma media and the photon energy $\hbar \omega >T_{e}$ is
described by dimensionless intensity parameter of intensity $\chi
_{0}=eE_{0}/\omega \sqrt{m_{e}\hbar \omega }$ \cite{book} ($E_{0}$ is the
wave field strength amplitude, $\omega $ is the frequency of the wave, $e$
and $m_{e}$ are the elementary charge and mass, $T_{e}$ is the electrons
temperature). The $\chi _{0}$ is the ratio of the amplitude of the momentum
given by the wave field to momentum at the one-photon absorption. Hence the
intensity of the wave expressed by the parameter $\chi _{0}$ can be
estimated as $I_{\chi _{0}}=\chi _{0}^{2}\times 1.74\times 10^{12}$ $\mathrm{%
Wcm}^{-2}(\hbar \omega /\mathrm{eV})^{3}$ \cite{7aaa}. Multiphoton effects
become essential at $\chi _{0}\sim 1$, which for terahertz photons $\hbar
\omega \sim 0.01$\ $\mathrm{eV}$ corresponds to intensity $I_{\chi
_{0}}=1.74\times 10^{6}$\ $\mathrm{Wcm}^{-2}$. Meanwhile, in BG for
intraband transitions the wave-particle interaction at the photon energies $%
\hbar \omega >\mathcal{E}_{L}$\ \cite{Mer4} characterizes by known
dimensionless intensity parameter $\chi =\chi _{0}(m_{e}\rightarrow m_{\ast
})$ ($m_{\ast }$ is an effective mass of chiral particle, $\mathcal{E}%
_{L}\simeq 1$ $\mathrm{meV}$ is the Lifshitz energy). Depending on the value
of the parameter $\chi $, three regimes of the wave-particle interaction can
take place: $\chi \ll 1$\ corresponds to one-photon interaction, $\chi \gg 1$%
, which is the static field limit or so called Schwinger regime \cite{12},
and $\chi \succsim 1$\ -- corresponds the multiphoton interaction \cite{7a}.
The wave intensity for the given $\chi $ can be expressed as $I_{\chi }=\chi
^{2}\times 6\times 10^{10}$ $\mathrm{Wcm}^{-2}(\hbar \omega /\mathrm{eV}%
)^{3} $ \cite{Mer4}. Comparison of this intensity threshold with the
analogous one for the plasma electrons shows the substantial difference $%
I_{\chi _{0}=1}/I_{\chi =1}\sim 30$ between the values of these thresholds
(instead of the light speed here stands up the Fermi velocity -much less
than the light speed in vacuum). Thus, for realization of multiphoton SB in
BG one can expect $\sim 30$\ times smaller intensities than for SB in atoms 
\cite{7aa}, \cite{7aaa}, \cite{12a,12aa,13}. In addition, the considered in
the present paper pump wave photon energy range of interest lies in the
terahertz or near infrared domain, where the high-power terahertz or
near-infrared generators and frequency multipliers are of special interest.
In general, the role of terahertz or near infrared radiation for the study
of nonlinear phenomena in condensed matter physics is important.

Regarding the investigation of electrons elastic scattering on impurity ions
in intrinsic SG, it has been investigated mainly within the framework of
perturbation theory by electrostatic potential (see, e.g., \cite%
{BornEllastic,Chen2008,BornEl1,BornEl2,BornEl3,TanAdam2007,Katsnelson}). The
SB process in graphene at moderate intensities of stimulated radiation, in
case of its linear absorption by electrons (or holes), there are extensive
investigations carried out in the scope of the linear theory (see, e.g. \cite%
{7b,7b1,7b2,7b3,Zhu2014}). Multiphoton cross-sections for SB of conductive
electrons in intrinsic SG have been obtained in the Born approximation over
the scattering potential in the presence of an external EM radiation (see, 
\cite{nanop1}, \cite{nanop2}). The transport properties of the BG are
studied theoretically within a self-consistent Born approximation in the
papers \cite{KoshinoAndo}. The transport and elastic scattering times on
impurities in SG and BG are investigated experimentally and comparison with
theoretical predictions have been made in \cite%
{Adam,Katsnel,ElastBilayer,Zhang,Sharma,Xiao}.

In the present paper the formerly developed \cite{nanop1} quantum mechanical
consideration of nonlinear stimulated scattering of charged carriers on the
Coulomb field of impurity ions in SG in the presence of an external strong
coherent EM radiation was extended to doped BG. The selected frequency range
of terahertz radiation excludes the valence electrons excitations at high
Fermi energies. The interaction of electrons with the scattering potential
is considered in the Born approximation. The analytic formulas in the case
of screened Coulomb potential have been analyzed numerically for actual
parameters of interaction system in the strong radiation field. The results
of our investigations show that in the terahertz domain of wave frequencies
one can achieve efficient rates of absorption of pump waves already at
moderate intensities by this mechanism.

In Sec. II the multiphoton rates and total cross-sections of SB process of
conductive electrons on the charged impurity ions in BG have been obtained.
In Sec. III the analytic formulas in case of screened Coulomb potential have
been analyzed numerically and the results for the partial differential
cross-sections of SB, as well as the angular dependence of scattered
electrons on the incident radiation intensity are received. Conclusions are
given in Sec. IV.

\section{Multiphoton transition amplitudes and cross-sections of 2D Dirac
fermions stimulated bremsstrahlung in doped bilayer graphene}

In the following we will develop the scattering theory for the 2D chiral
fermions on arbitrary electrostatic potential $U(\mathbf{r})$ of an impurity
ion in doped BG at the presence of external EM\ wave of moderate
intensities, in analogous with intrinsic SG \cite{nanop1}. Using the
scattering Green function formalism in the Born approximation by potential $%
U(\mathbf{r})$ we will obtain the SB rates. Note, that the pioneer treatment
of nonrelativistic SB in the Born approximation has been carried out
analytically in \cite{12a}. Then this approach has been extended to the
relativistic domain \cite{12aa}, \cite{13}. In accordance with the quantum
theory, the transition amplitude of SB process in the coherent EM radiation
field from the state with the canonical momentum $\mathbf{p}%
_{0}(p_{x},p_{y}) $ to the state with momentum $\mathbf{p}$ $(p_{x},p_{y})$
in graphene plane ($X,Y$), can be written as:

\begin{equation}
T_{\mathbf{pp}_{0}}=-\frac{i}{\hbar }\int \Psi _{\mathbf{p}}^{+}(\mathbf{r}%
,t)U(\mathbf{r})\Psi _{\mathbf{p}_{0}}(\mathbf{r},t)dtd\mathbf{r,}  \label{1}
\end{equation}%
where bispinor function $\Psi ^{+}$ is the complex conjugation of $\Psi $.

We use an even simpler model for BG which neglects both the electron-hole
asymmetry and the trigonal warping. To exclude the valence electrons
excitations at high Fermi energies in graphene, we will assume for EM\ wave
to be a terahertz radiation. The BG system is still chiral due to the $A/B$
sublattice symmetry giving rise to the conserved pseudospin quantum index.
The detailed chiral $4$-component wavefunction for the bilayer case
including both layer and sublattice degrees of freedom can be found in the
original literature in \cite{2a}, \cite{Screening}, \cite{6c}, \cite%
{Nilsson1}, \cite{Nilsson2}. In accordance with the nonlinear quantum theory
of BG, in the mentioned case with energy window $0.002$ $\mathrm{eV}\lesssim
\varepsilon \lesssim 0.1$ $\mathrm{eV}$ in the vicinity of Dirac points $%
K_{\zeta }$ (valley quantum number $\zeta =\pm 1$) in the Brillouin zone,
the fermion particle wave function $\Psi _{\mathbf{p}}(\mathbf{r},t)$ in the
strong EM wave field may be presented in the form: 
\begin{equation}
\Psi _{\mathbf{p}}(\mathbf{r},t)=\frac{1}{\sqrt{S}}F_{\mathbf{p}\sigma }e^{%
\frac{i}{\hbar }\mathbf{pr}-i\Omega (\mathbf{p},t)}.  \label{2}
\end{equation}%
Here parameter $S$ is the quantization area--graphene layer surface area, 
\begin{equation}
F_{\mathbf{p}\sigma }=\frac{1}{\sqrt{2}}\left( 
\begin{array}{c}
e^{i2\zeta \vartheta } \\ 
\sigma%
\end{array}%
\right) ,  \label{3}
\end{equation}%
\begin{equation}
\Omega (\mathbf{p},t)=\frac{1}{2m_{\ast }\hbar }\int \left[ \left( p_{x}+%
\frac{e}{c}A_{x}\right) ^{2}+p_{y}^{2}\right] dt,  \label{3a}
\end{equation}%
are the spinor wave function $F_{\mathbf{p}\sigma }$ and the temporal phase $%
\Omega (\mathbf{p},t)$. The function $\vartheta (\mathbf{p}+\frac{e}{c}%
\mathbf{A}(t))$, given in the form: 
\begin{equation}
\vartheta (\mathbf{p}+\frac{e}{c}\mathbf{A}(t))=\arctan \left( p_{y}/\left(
p_{x}+\frac{e}{c}A_{x}\right) \right) ,  \label{3aa}
\end{equation}%
defines the angle between the vectors of particle kinematic momentum $%
\mathbf{p}=\mathbf{p}\left( p_{x}+\frac{e}{c}A_{x},p_{y}\right) $ in the EM
field and the wave vector potential $\mathbf{A}(t)=-c\int_{0}^{t}\mathbf{E}%
(t^{\prime })dt^{\prime }$ (unit vector $\widehat{\mathbf{e}}$ is directed
along the axis $OX$). The BG has a low-energy dispersion which is
approximated \cite{3b}, \cite{Nilsson2} by massive valence and conduction
bands without gap (in opposite to zero-mass bands in SG). The quasiparticle
energy $\varepsilon (p)$ is defined by $\varepsilon (p)=\sigma \left(
p_{x}^{2}+p_{y}^{2}\right) /2m_{\ast }$ , where $\sigma =\pm 1$ correspond
to the conduction/valence bands; $m_{\ast }=\gamma _{1}/\left( 2\mathrm{v}%
_{F}^{2}\right) $, $\gamma _{1}\simeq 0.39$ $\mathrm{eV}$ is the interlayer
tunneling amplitude, $\mathrm{v}_{F}$ is the intrinsic SG Fermi velocity.
This parabolic dispersion applies only for small values of $p$ satisfying $%
p/\hbar \ll \gamma _{1}/\left( \hbar \mathrm{v}_{F}\right) $ and for carrier
densities (or equivalently, the impurity densities) smaller than $5\times
10^{12}\mathrm{cm}^{-2}$, respectively. In the opposite limit, $p/\hbar \gg
\gamma _{1}/\left( \hbar \mathrm{v}_{F}\right) $ and for the carrier
densities larger than the last, we get a linear band dispersion $\varepsilon
(p)=\sigma \hbar p\mathrm{v}_{F}$, just as in the intrinsic SG case \cite%
{Ando1982}. For the actual parameters the effective mass for BG is $m_{\ast
}\approx \left( 0.033\div 0.05\right) m_{e}$, with free electron mass $m_{e}$%
.

The spin and the valley quantum numbers are conserved. There is no
degeneracy upon the valley quantum number $\zeta $ for the issue considered.
However, since there are no intervalley transitions, the valley index can be
considered as a parameter.

The wave will be applied in the perpendicular direction to the BG sheet ($XY$%
) to exclude the effect of the magnetic field. For the simplest the plane
quasimonochromatic EM wave of frequency $\omega $ is taken to be linearly
polarized along the $X$ axis: $E(t)=$ $E_{0}\cos \omega t$. It is clear that
at such small Fermi velocities of scattered particles the plane
monochromatic wave will turn into the uniform periodic electric field of
frequency $\omega =2\pi /T$. Similar calculations for a wave linearly
polarized along the $X$ axis show qualitatively the same picture.

The known formalism of the time dependent scattering theory can be applied
to obtain the multiphoton ($\chi \succsim 1$) SB process transition
amplitude $T_{\mathbf{pp}_{0}}$ between the unperturbed states $\Psi _{%
\mathbf{p}}$ and $\Psi _{\mathbf{p}_{0}}$. Thus, we can write for transition
amplitude the following ratio:

\begin{equation}
T_{\mathbf{pp}_{0}}=-\frac{i}{\hbar }\int F_{\mathbf{p}}^{+}(t)U(\mathbf{r}%
)F_{\mathbf{p}_{0}}(t)\exp \left( \frac{i}{\hbar }\left( \mathbf{p-p}%
_{0}\right) \mathbf{r}\right) dtd\mathbf{r,}  \label{4}
\end{equation}%
which can be expressed in the form:%
\begin{equation}
T_{\mathbf{pp}_{0}}=-\frac{i}{\hbar }\int F_{\mathbf{p}}^{+}(t)\widetilde{U}%
\left( \frac{\mathbf{p-p}_{0}}{\hbar }\right) F_{\mathbf{p}_{0}}(t)dt,
\label{5}
\end{equation}%
where 
\begin{equation}
\widetilde{U}\left( \frac{\mathbf{p-p}_{0}}{\hbar }\right) =\int \exp \left( 
\frac{i}{\hbar }\left( \mathbf{p-p}_{0}\right) \mathbf{r}\right) U(\mathbf{r}%
)d\mathbf{r}  \label{6}
\end{equation}%
is the Fourier transform of the scattering potential. Taking into account
the relation (\ref{5}) for the impurity potential of the arbitrary form $%
\widetilde{U}$, we will have for $T_{\mathbf{pp}_{0}}$ the relation:

\begin{equation*}
T_{\mathbf{pp}_{0}}=-\frac{i}{\hbar }\frac{1}{2S}\widetilde{U}\left[ \frac{%
\mathbf{p-p}_{0}}{\hbar }\right]
\end{equation*}%
\begin{equation}
\times \int \left( 1+e^{i2\zeta \left[ \vartheta (\mathbf{p}_{0}+\frac{e}{c}%
\mathbf{A}(\tau ))-\vartheta (\mathbf{p}+\frac{e}{c}\mathbf{A}(t))\right]
}\right)  \label{7}
\end{equation}%
\begin{equation*}
\times e^{-\frac{i}{\hbar }\frac{1}{2m_{\ast }}\int_{0}^{\tau }\left[ \left(
p_{x}+\frac{e}{c}A_{x}\right) ^{2}+p_{y}^{2}-\left( p_{0x}+\frac{e}{c}%
A_{x}\right) ^{2}+p_{0y}^{2}\right] dt}d\tau .
\end{equation*}%
Using the Eq. (\ref{7}), the impurity ion potential can be represented in
the following formula:%
\begin{equation}
T_{\mathbf{pp}_{0}}=\int B(\tau )e^{-\frac{i}{\hbar }\left( \mathcal{E}-%
\mathcal{E}_{0}\right) \tau }d\tau ,  \label{15}
\end{equation}%
where the periodic function $B(\tau )$ of time defines as 
\begin{equation*}
B(\tau )=-\frac{i}{\hbar }\frac{1}{2S}\widetilde{U}\left[ \frac{\mathbf{p-p}%
_{0}}{\hbar }\right]
\end{equation*}%
\begin{equation*}
\times \left( 1+e^{i2\zeta \left[ \vartheta (\mathbf{p}_{0}+\frac{e}{c}%
\mathbf{A}(t))-\vartheta (\mathbf{p}+\frac{e}{c}\mathbf{A}(t))\right]
}\right)
\end{equation*}%
\begin{equation}
\times e^{-\frac{i}{\hbar }\int_{0}^{\tau }\left[ \left( \frac{1}{2m_{\ast }}%
\left( p_{x}+\frac{e}{c}A_{x}\right) ^{2}+\frac{p_{y}^{2}}{2m_{\ast }}-%
\mathcal{E}\right) -\left( \frac{1}{2m_{\ast }}\left( p_{0x}+\frac{e}{c}%
A_{x}\right) ^{2}+\frac{p_{0y}^{2}}{2m_{\ast }}-\mathcal{E}_{0}\right) %
\right] dt},  \label{17}
\end{equation}%
and the time-depended modules of the "quasienergies" $\mathcal{E}_{0},%
\mathcal{E}$\ are:%
\begin{equation*}
\mathcal{E}_{0}=\frac{\omega }{4\pi m_{\ast }}\int_{0}^{\frac{2\pi }{\omega }%
}\left[ \left( p_{0x}+\frac{e}{c}A_{x}(t)\right) ^{2}+p_{0y}^{2}\right] dt,
\end{equation*}%
\begin{equation*}
\mathcal{E}=\frac{\omega }{4\pi m_{\ast }}\int_{0}^{\frac{2\pi }{\omega }}%
\left[ \left( p_{x}+\frac{e}{c}A_{x}(t)\right) ^{2}+p_{y}^{2}\right] dt,
\end{equation*}%
(with corresponding "quasimoumentums" $\mathbf{P}_{0}\left\{ p_{0x}+\frac{e}{%
c}A_{x}(t),p_{0y}\right\} ,\mathbf{P}\left\{ p_{x}+\frac{e}{c}%
A_{x}(t),p_{y}\right\} $). Further we made the Fourier transformation of the
function $B(t)$\ over $t$ by the relations:%
\begin{equation}
B(t)=\sum\limits_{n=-\infty }^{\infty }\widetilde{B}_{n}\exp (-in\omega t),
\label{19}
\end{equation}%
\begin{equation}
\widetilde{B}_{n}=\frac{\omega }{2\pi }\int_{0}^{2\pi /\omega }B(t)\exp
(in\omega t)dt.  \label{20}
\end{equation}%
Thus, carried out the integration over $t$ in the formula (\ref{15}), we
have obtained for the transition amplitude $T_{\mathbf{pp}_{0}}$:%
\begin{equation}
T_{\mathbf{pp}_{0}}=2\pi \hbar \widetilde{B}_{n}\delta \left( \mathcal{E}-%
\mathcal{E}_{0}-n\hbar \omega \right) .  \label{21}
\end{equation}

The differential probability $dW_{\mathbf{pp}_{0}}$ of SB process in strong
pump wave field per unit time, from the electron or hole state with
two-dimensional momentum $\mathbf{p}_{0}$ to a state with momentum $\mathbf{p%
}$ in the phase space $g_{v}g_{s}Sd\mathbf{P/}(2\pi \hbar )^{2}$ \cite%
{Ando1982} in the Born approximation may be described by the formula:

\begin{equation}
dW_{\mathbf{pp}_{0}}=\underset{t\rightarrow \infty }{\lim }\frac{1}{t}%
\left\vert T_{\mathbf{pp}_{0}}\right\vert ^{2}g_{v}g_{s}S\frac{d\theta PdP}{%
(2\pi \hbar )^{2}},  \label{22}
\end{equation}%
where $d\theta $ is the differential scattering angle. Here the degeneracy
factor is $g_{v}g_{s}=4$ due to valley and spin degeneracies.

Dividing the differential probability $W_{\mathbf{pp}_{0}}$ (\ref{21}) of SB
by initial flux density $\left\vert \mathbf{P}_{0}\right\vert /m_{\ast }$
and integrating over $dP$, we can obtain the multiphoton differential
cross-section $d\Lambda /d\theta $ of SB for quasiparticles in doped BG:

\begin{equation}
\frac{d\Lambda }{d\theta }=\underset{n=-n_{\max }}{\sum^{\infty }}\frac{%
d\Lambda ^{\left( n\right) }}{d\theta },  \label{23}
\end{equation}%
where

\begin{equation}
\frac{d\Lambda ^{\left( n\right) }}{d\theta }=\frac{2m_{\ast }^{2}}{\pi
\hbar ^{2}}\left\vert \frac{\mathbf{P}}{\mathbf{P}_{0}}\right\vert \left.
\left\vert \widetilde{B}_{n}\right\vert ^{2}\right\vert
_{P^{2}=P_{0}^{2}+2m_{\ast }\hbar \omega n}  \label{24}
\end{equation}%
is the partial differential cross-section of $n$-photon SB\ with maximum
number of emitted photons $n_{\max }$. The total multiphoton scattering
cross-section $d\Lambda /d\theta $\ will be obtained making summation over
the photon numbers in the formula for differential partial cross-sections $%
d\Lambda ^{\left( n\right) }/d\theta $ (\ref{23}), which may be expressed by
the following form:

\begin{equation*}
\frac{d\Lambda ^{\left( n\right) }}{d\theta }=\left\vert \widetilde{U}\left[ 
\frac{\mathbf{p-p}_{0}}{\hbar }\right] \right\vert ^{2}
\end{equation*}%
\begin{equation*}
\times \left\vert \int_{0}^{T}d\left( \frac{t}{T}\right) \left( 1+e^{i2\zeta %
\left[ \vartheta (\mathbf{p}_{0}+\frac{e}{c}\mathbf{A}(t))-\vartheta (%
\mathbf{p}+\frac{e}{c}\mathbf{A}(t))\right] }\right) \exp (in\omega t)\right.
\end{equation*}%
\begin{equation*}
\times \left. e^{-\frac{i}{\hbar }\int_{0}^{\tau }\left[ \left( \frac{1}{%
2m_{\ast }}\left( p_{x}+\frac{e}{c}A_{x}\right) ^{2}+\frac{p_{y}^{2}}{%
2m_{\ast }}-\mathcal{E}\right) -\left( \frac{1}{2m_{\ast }}\left( p_{0x}+%
\frac{e}{c}A_{x}\right) ^{2}+\frac{p_{0y}^{2}}{2m_{\ast }}-\mathcal{E}%
_{0}\right) \right] dt^{\prime }}\right\vert ^{2}
\end{equation*}%
\begin{equation}
\times \delta \left( \frac{P^{2}}{2m_{\ast }}-\frac{P_{0}^{2}}{2m_{\ast }}%
-\hbar \omega n\right) \frac{2m_{\ast }PdP}{P_{0}\pi \hbar ^{3}}.
\label{exact1}
\end{equation}%
We produce the expansion in Eq. (\ref{exact1}) into a Taylor series, keeping
only the terms of the first order over the electric field and we will obtain
the partial differential cross-sections of SB in the case of linear theory.
Thus, we can obtain the asymptotic formula for the partial differential
cross-sections of SB process in the weak wave field $\mathbf{A}(t)$:

\begin{equation*}
\frac{d\Lambda ^{\left( \mathrm{linear}\right) }}{d\theta }=\frac{2m_{\ast
}^{2}}{\pi \hbar ^{3}\left\vert \mathbf{p}_{0}\right\vert }\left\vert 
\widetilde{U}\left[ \frac{\mathbf{p-p}_{0}}{\hbar }\right] \right\vert
^{2}\left\vert \left( 1+e^{i2\zeta \left[ \vartheta (\mathbf{p}%
_{0})-\vartheta (\mathbf{p}))\right] }\right) \delta _{n0}\delta _{\mathcal{%
EE}_{0}}\right.
\end{equation*}%
\begin{equation*}
+\left[ \frac{\chi }{2i}C_{1}\delta _{n-1}\right] \delta _{\mathcal{E}\left( 
\mathcal{E}_{0}+\hbar \omega \right) }
\end{equation*}%
\begin{equation}
\left. -\left[ \frac{\chi }{2i}C_{1}\delta _{n+1}\right] \delta _{\mathcal{E}%
\left( \mathcal{E}_{0}-\hbar \omega \right) }\right\vert ^{2}d\mathcal{E}.
\label{aaa}
\end{equation}%
Here the time dependent $C_{1}$coefficient is defined as:%
\begin{equation*}
C_{1}=-\left( \frac{p_{x}}{\sqrt{m_{\ast }\hbar \omega }}-\frac{p_{0x}}{%
\sqrt{m_{\ast }\hbar \omega }}\right) \left( 1+e^{i2\zeta \left[ \vartheta (%
\mathbf{p}_{0})-\vartheta (\mathbf{p}))\right] }\right)
\end{equation*}%
\begin{equation}
+2\zeta \frac{\left( \mathcal{E-E}_{0}\right) }{\hbar \omega }\left[ \frac{%
\sqrt{m_{\ast }\hbar \omega }p_{y}}{p^{2}}-\frac{\sqrt{m_{\ast }\hbar \omega 
}p_{0y}}{p_{0}^{2}}\right] e^{i2\zeta \left[ \vartheta (\mathbf{p}%
_{0})-\vartheta (\mathbf{p})\right] }.  \label{aaa1}
\end{equation}%
Comparing the $n$-photon cross-section $d\Lambda ^{\left( n\right) }$ (\ref%
{exact1}) of SB process with the elastic one, we conclude that formula (\ref%
{exact1}) at $\mathbf{A}(\tau )=0$ ($n=0$) passes to elastic scattering
cross-section $d\Lambda _{elast}$ \cite{KoshinoAndo}, which is\ the analog
of the Mott formula but in 2D scattering theory:%
\begin{equation*}
\frac{d\Lambda _{elast}}{d\theta }=\frac{2m_{\ast }^{2}}{\pi \hbar
^{3}\left\vert \mathbf{p}_{0}\right\vert }\left\vert \widetilde{U}\left[ 
\frac{\mathbf{p-p}_{0}}{\hbar }\right] \right\vert ^{2}
\end{equation*}%
\begin{equation}
\times \left\vert \left( 1+e^{i2\zeta \left[ \vartheta (\mathbf{p}%
_{0})-\vartheta (\mathbf{p})\right] }\right) \right\vert ^{2}.
\label{exact3}
\end{equation}%
The phase term 
\begin{equation*}
\left( 1+\exp i2\zeta \left[ \vartheta (\mathbf{p}_{0}+\frac{e}{c}\mathbf{A}%
(\tau )-\vartheta (\mathbf{p}+\frac{e}{c}\mathbf{A}(\tau ))\right] \right)
^{2}
\end{equation*}%
in Eq. (\ref{exact1}) at $\mathbf{A}(\tau )=0$ is the known overlap factor 
\begin{equation}
\left( 1+\sigma e^{-i\theta _{q}}\right) \left( 1+\sigma ^{\prime
}e^{i\theta _{q}}\right) =2\left( 1+\sigma \sigma ^{\prime }\cos \theta
_{q}\right) ,  \label{333}
\end{equation}%
where $\theta _{q}=2\zeta \vartheta (\mathbf{p}_{0})-2\zeta \vartheta (%
\mathbf{p})$. The term (\ref{333}) is a wave function form-factor associated
with the chiral nature of BG, and the last is determined by its band
structure.

\section{Cross-sections of stimulated bremsstrahlung on the screening
Coulomb potential of impurity ions in doped bilayer graphene}

Bellow we will obtain the partial differential cross-sections in particular
case of SB process on a screened Coulomb potential of impurity ions in BG,
using Eq. (\ref{exact1}). In an analytic form of Coulomb screening in BG 
\cite{Screening}, \cite{Ando1982,Sarma2011,Xu2} has been performed in
analogous with the case of intrinsic SG \cite{BornEl2}, \cite{Katsnelson}.
So, the Fourier transform $\widetilde{U}\left( \mathbf{q}\right) =\int U(%
\mathbf{r})e^{-i\mathbf{qr}}d\mathbf{r}$ of a charged impurity center of
potential\textbf{\ }can be written as:%
\begin{equation}
\widetilde{U}\left( \mathbf{q}\right) =\frac{2\pi e^{2}}{\kappa q\epsilon
\left( q\right) },  \label{29}
\end{equation}%
where the screening term $\epsilon \left( q\right) $ ($q=\left\vert \mathbf{q%
}\right\vert $) is the 2D finite temperature static dielectric function in
random phase approximation (RPA) appropriate for BG, given by the formula 
\cite{Screening}, \cite{Sarma2011}:%
\begin{equation}
\epsilon \left( q\right) =1+\frac{q_{s}}{q}\left[ g\left( q\right) -f\left(
q\right) \Theta \left( q-2k_{F}\right) \right] .  \label{30}
\end{equation}%
Here $k_{F}=\sqrt{2m_{\ast }\varepsilon _{F}}/\hbar $ is 2D Fermi wave
vector in BG case, $q_{s}=k_{TF}/k_{F}=4m_{\ast }e^{2}\log 4/\left( \kappa
\hbar ^{2}\right) $ ($\sim n^{-1/2}$ for BG \cite{Sarma2011}) is the 2D
Thomas-Fermi screening wave vector given by $k_{TF}$ \cite{Ando1982} scaled
on $k_{F}$, which controls the dimensionless strength of quantum screening;
and $\kappa $ is the background lattice dielectric constant of the system.
The function $\Theta \left( q-2k_{F}\right) $ is the step function (or
staircase function). The functions $g\left( q\right) $, $f\left( q\right) $
are defined by the formulas \cite{Sarma2011}:%
\begin{equation}
g\left( q\right) =\frac{1}{2}\sqrt{4+\overline{q}^{4}}-\log \left[ \frac{1+%
\sqrt{1+\overline{q}^{4}/4}}{2}\right] ,  \label{31}
\end{equation}%
\begin{equation}
f\left( q\right) =\left( 1+\frac{\overline{q}^{2}}{2}\right) \sqrt{1-\frac{4%
}{\overline{q}^{2}}}+\log \left[ \frac{\overline{q}-\sqrt{\overline{q}^{2}-4}%
}{\overline{q}+\sqrt{\overline{q}^{2}-4}}\right] ,  \label{31a}
\end{equation}%
where $\overline{q}=q/k_{F}$. This usual 2D dispersion or the static
screening \cite{Screening} one is the particular case of the wave vector
dependent plasmon dispersion and the wave frequency dependent screening
function case \cite{Sarma2007}. For the last dielectric function and
screening show very different behavior than in usual 2D graphene case, and
the plasmons creation will be actual. 
\begin{figure}[tbp]
\includegraphics[width=.51\textwidth]{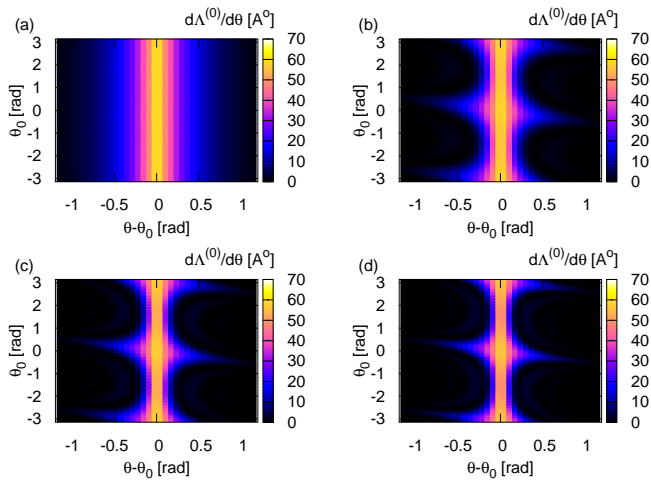}
\caption{{}(Color online) Partial differential cross section $d\Lambda
^{(n)}/d\protect\theta $\ (in angstrom) of SB process for photon number $n=0$
vs the electron deflection angle $\protect\theta -\protect\theta _{0}$\ and
incident angle $\protect\theta _{0}$ for linear polarization of EM wave of
intensities: (a) $\protect\chi =0$, (b) $\protect\chi =1$, (c) $\protect\chi %
=1.5$, and (d) $\protect\chi =2$.}
\end{figure}
\begin{figure}[tbp]
\includegraphics[width=.51\textwidth]{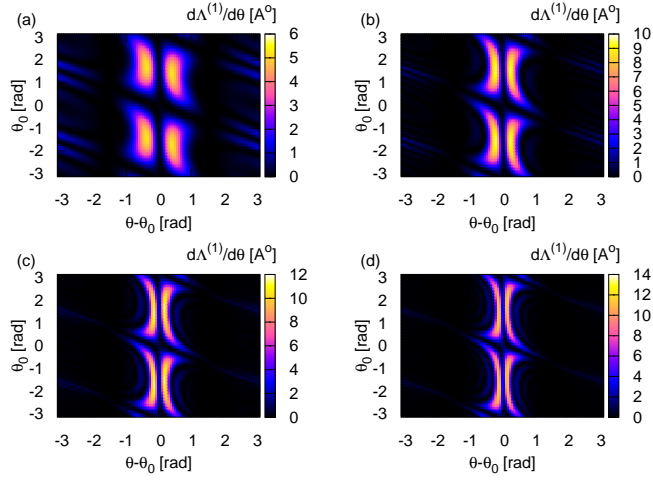}
\caption{ (Color online) Partial differential cross section $d\Lambda
^{(n)}/d\protect\theta $\ (in angstrom) of SB process for photon number $n=1$
and laser intensities: (a) $\protect\chi =0.5$, (b) $\protect\chi =1$, (c) $%
\protect\chi =1.5$, and (d) $\protect\chi =2$.}
\end{figure}

For the following numerical considerations it is convenient to represent the
differential cross-sections of SB on the charged impurities in the form of
dimensionless quantities. We will obtain the final form for the
dimensionless partial differential cross-sections of SB process $d\Lambda
^{\left( n\right) }/d\theta $, taking into account Eqs. (\ref{7}), (\ref%
{exact1}), and (\ref{29}) with relations Eqs. (\ref{30})--(\ref{31a}), in
the certain case of linearly polarized EM\ wave with the dimensionless
vector potential $\overline{\mathbf{A}}(t)=-\widehat{\mathbf{e}}{\chi }\sin
(2\pi \tau )$ scaled on the wavelength $\lambda =2\pi c/\omega $ (unit
vector $\widehat{\mathbf{e}}$ is directed along the axis $OX$):%
\begin{equation*}
\frac{d\Lambda ^{\left( n\right) }}{\lambda d\theta }=\frac{1}{2400\log ^{2}4%
}\frac{1}{\overline{\varepsilon }_{0}}\frac{\left\vert \overline{\mathbf{p}}%
_{0}\right\vert }{\overline{m_{\ast }\mathrm{v}_{F}}}
\end{equation*}%
\begin{equation*}
\times \frac{\overline{\hbar q}_{s}^{2}}{\left\vert \overline{\mathbf{p}}%
\mathbf{-}\overline{\mathbf{p}}_{0}\right\vert ^{2}\epsilon ^{2}\left(
\left\vert \overline{\mathbf{p}}\mathbf{-}\overline{\mathbf{p}}%
_{0}\right\vert \right) }
\end{equation*}%
\begin{equation*}
\times \left\vert \int_{0}^{1}d\tau \left( 1+e^{i2\zeta \left[ \vartheta (%
\overline{\mathbf{p}}_{0}-{\chi }\sin (2\pi \tau ))-\vartheta (\overline{%
\mathbf{p}}-{\chi }\sin (2\pi \tau ))\right] }\right) \right.
\end{equation*}%
\begin{equation*}
\times \exp \left\{ i2\pi n\tau -\pi i\int_{0}^{\tau }\left[ \left( \left( 
\overline{p}_{x}-{\chi }\sin (2\pi \tau ^{\prime })\right) ^{2}+\overline{p}%
_{y}^{2}-2\overline{\mathcal{E}}\right) \right. \right.
\end{equation*}%
\begin{equation*}
\left. \left. \left. -\left( \left( \overline{p}_{0x}-{\chi }\sin (2\pi \tau
^{\prime })\right) ^{2}+\overline{p}_{0y}^{2}-2\overline{\mathcal{E}}%
_{0}\right) \right] d\tau ^{\prime }\right\} \right\vert ^{2}
\end{equation*}%
\begin{equation}
\times \delta \left( \overline{\mathcal{E}}-\overline{\mathcal{E}}%
_{0}-n\right) d\overline{\mathcal{E}}.  \label{exact2}
\end{equation}%
In Eq. (\ref{exact2}) the dimensionless momentum, energy, 2D screening
vector, 2D Fermi velocity, and time are defined as follows:%
\begin{equation}
\overline{p}_{x,y}=\frac{p_{x,y}}{\sqrt{m_{\ast }\hbar \omega }},\overline{%
\mathcal{E}}=\frac{\mathcal{E}}{\hbar \omega },  \label{32}
\end{equation}

\begin{equation}
\overline{\hbar q}_{s}=\frac{\hbar q_{s}}{\sqrt{m_{\ast }\hbar \omega }},%
\overline{m_{\ast }\mathrm{v}_{F}}=\frac{m_{\ast }\mathrm{v}_{F}}{\sqrt{%
m_{\ast }\hbar \omega }},d\tau =\frac{dt}{T}.  \label{33a}
\end{equation}%
\begin{figure}[tbp]
\includegraphics[width=.51\textwidth]{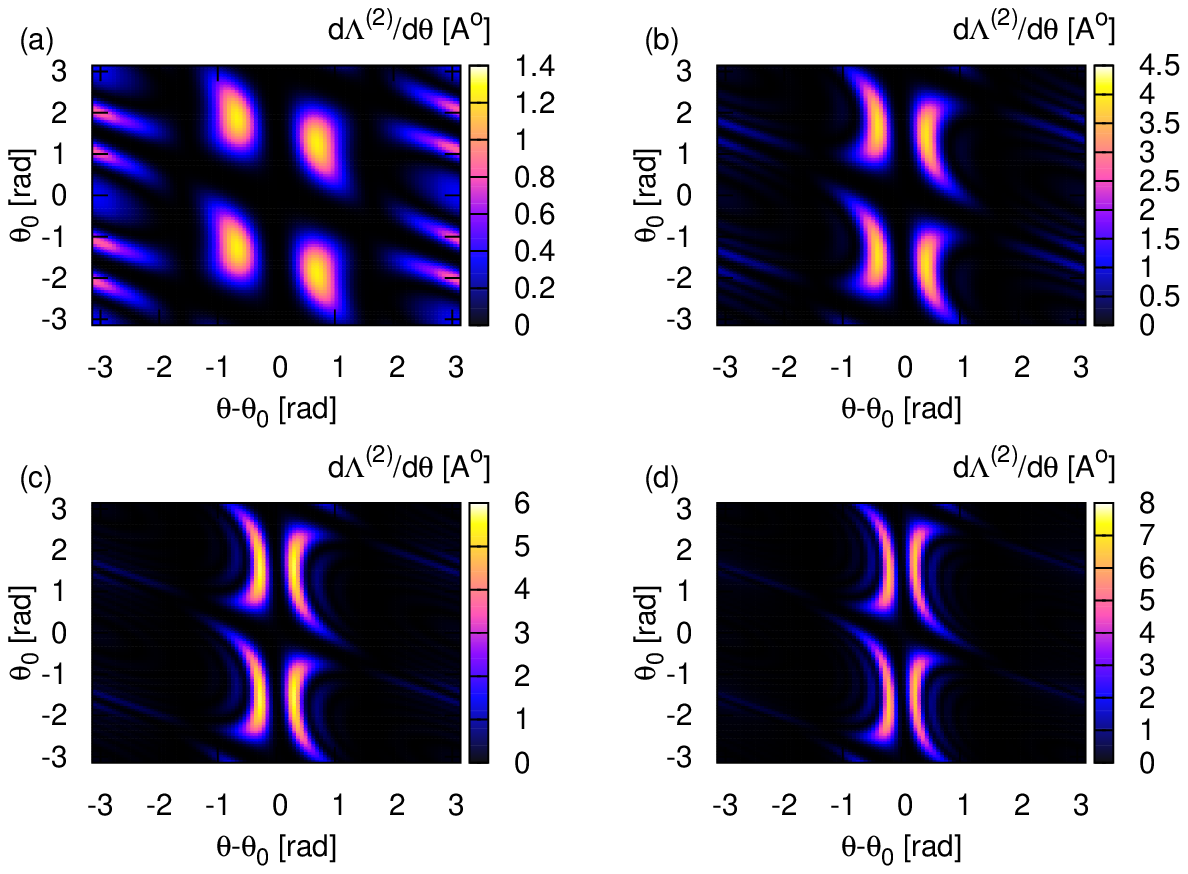}
\caption{ (Color online) Partial differential cross section $d\Lambda
^{(n)}/d\protect\theta $\ (in angstrom) for photon number $n=2$: ($a$)--($d$%
) correspond to dimensionless field parameters $\protect\chi =0.5,1,1.5$ and 
$2$, respectively.}
\end{figure}
\begin{figure}[tbp]
\includegraphics[width=.51\textwidth]{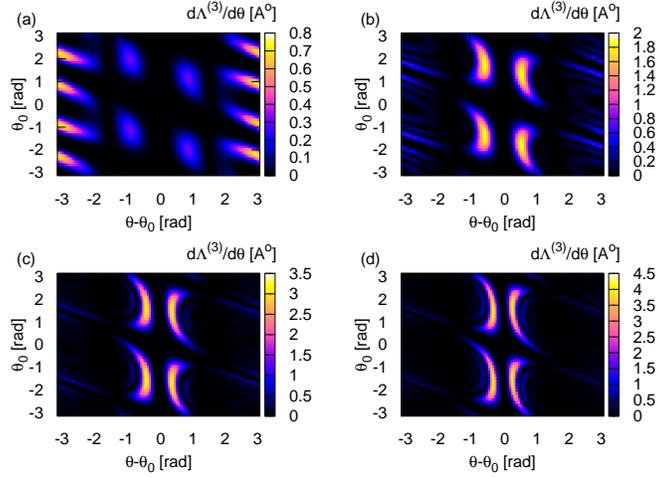}
\caption{(Color online) Partial differential cross section $d\Lambda
^{(n)}/d\protect\theta $\ (in angstrom) for photon number $n=3$: ($a$)--($d$%
) correspond to dimensionless field parameters $\protect\chi =0.5,1,1.5$ and 
$2$, respectively.}
\end{figure}

As mentioned in Introduction, the intensity $I_{\chi }$ strongly depends on
the parameter $\chi $ and the photon energy. At $\chi \gtrsim 1$ the
multiphoton effects become essential. So, for numerical analysis of
multiphoton SB\ cross sections in graphene we assume coherent EM radiation
with intensity dimensionless parameter $\chi \gtrsim 1$ in particular case
of the fixed photon energy $\hbar \omega =0.01$ \textrm{eV }($\lambda
=1.24\times 10^{6}\mathrm{\mathring{A}}$) from terahertz domain. The Fermi
energy is taken to be $\varepsilon _{F}=20\hbar \omega $ ($\varepsilon
_{F}\gg \hbar \omega $), dielectric environment constant is $\kappa =2.5$
for an impurity strength in the presence of the $\mathrm{SiO}_{2}$ substrate 
\cite{BLG}. Note, that while for graphene on $\mathrm{SiO}_{2}$ substrate
corresponding to $\kappa =2.5$ \cite{Katsnelson}, in particularly, covering
BG on $\mathrm{SiO}_{2}$ with ice, it can be change the last to $\kappa =3.5$%
, which would increase the conductivity by $10\%$ for permitted charged
impurity density $\sim 10^{12}$ $\mathrm{cm}^{-2}$ \cite{BLG}. The band
dispersion is quadratic and only the lowest subband is occupied \cite%
{2e,3a,Screening}, and the RPA theory applies in the density domain $10^{10}$
$\mathrm{cm}^{-2}<n<5\times 10^{12}$ $\mathrm{cm}^{-2}$. In particular case
for a homogeneous carrier density $n=10^{11}$ $\mathrm{cm}^{-2}$, the
dimensionless effective screening vector $q_{s}$ defined by the formula \cite%
{Sarma2011}:%
\begin{equation*}
q_{s}=\frac{54.8}{\sqrt{n\times 10^{-10}\mathrm{cm}^{2}}},
\end{equation*}%
we have taken $q_{s}=17.329$. In the context of graphene, it has given
useful direct comparison between screening in SG versus screening in BG: 
\begin{equation*}
q_{TF}^{BG}/q_{TF}^{SG}\approx 16/\sqrt{n\times 10^{-10}\mathrm{cm}^{2}},
\end{equation*}%
which showing that as carrier density decreases, BG screening becomes much
stronger than SG screening \cite{Sarma2011}. The quantum dynamics of
multiphoton emission and absorption processes at the electron-impuarity ion
SB is studied making time integration in Eq. (\ref{exact2}) by the
fourth-order Runge-Kutta method. The all numerical results are for the
valley $\zeta =1$. The figures in the following illustrate the SB\ at
different intensities of stimulated wave of linear polarization. To reveal
the peculiarities which can be associated with the chiral nature of BG and
its parabolic dispersion, the comparison with the SG case has been made.

In the Figs. 1-5, the envelopes of partial differential cross sections $%
d\Lambda ^{(n)}/d\theta $ (\ref{exact2}) of SB in BG are plotted as a
function of the electron deflection angle $\theta -\theta _{0}$ and the
angle $\theta _{0}$ between the vector of the electron initial momentum and
electric field strengths of an coherent EM radiation field. Then, Fig. 1
shows the elastic scattering part ($n=0$), and Figs. 2-5 correspond to SB
for different number of absorbed photons $n>0$. Similar calculations for
emitted photons at $n<0$ show qualitatively the same picture, so that it is
omitted. The angular dependence of partial differential cross sections in
these figures are displayed for various intensities: in Fig. 1 (a) at $\chi
=0$, (b) $\chi =1$ ($I_{\chi }=6\times 10^{4}$ $\mathrm{Wcm}^{-2}$), (c) $%
\chi =1.5$ ($I_{\chi }=1.35\times 10^{5}$ $\mathrm{Wcm}^{-2}$), and (d) $%
\chi =2$ ($I_{\chi }=2.4\times 10^{5}$ $\mathrm{Wcm}^{-2}$), and in Figs.
2--5: (a) $\chi =0.5$ ($I_{\chi }=1.5\times 10^{4}$ $\mathrm{Wcm}^{-2}$),
(b) $\chi =1$, (c) $\chi =1.5$, and (d) $\chi =2$. As is seen, the maximum
values of the multiphoton SB partial cross sections at $n=0$ are reached in
the direction of the electric field at $\theta -\theta _{0}\simeq 0$. But
with the increase of the photon number and the wave intensity the angular
distribution becomes more asymmetric. In comparison with the SG case one can
note, that maximum values of the multiphoton cross sections correspond
substantially different values of the deflection angle $\theta -\theta _{0}$%
. Particularly, with the increase of the photon number, the multiphoton
cross section maximums at a weak wave intensity $\chi =0.5$ (closed to main
elastic result) are reached at deflection angles $\theta -\theta _{0}=\pm
\pi $. This demonstrates that the probability of the scattering under large
angles (backscattering) is considerable. As is seen from theses figures,
with the increase of the photon number and the wave intensity, the SB
partial differential cross sections' maximums are shifted to the direction
of the wave electric field.

To demonstrate the dependence of SB\ cross sections on the number of emitted
or absorbed photons, the envelopes of the partial cross sections $n\Lambda
^{(n)}$ for various angle $\theta _{0}$ between the initial electron
momentum and the wave electric field, are presented in Fig.6. For this goal,
the SB partial differential cross section (\ref{exact2}) is integrated over
the scattering angle of the outgoing electron for fixed initial angle $%
\theta _{0}$ and\ at various pump wave intensities. For terahertz photons,
the multiphoton interaction regime in BG can be achieved already at the
intensities $I_{\chi }\sim 10^{4}$ $\mathrm{Wcm}^{-2}$. Figure 6 shows that
with the increase of coherent radiation intensity, the multiphoton cut-off
number increases; and, compared to the one-photon scattering in linear
theory (\ref{aaa}), the multiphoton effect becomes dominant. As one can see,
at considered for numerical calculations intensities multiphoton SB process
have strongly nonlinear behavior. Hence, one can expect strong deviation for
the wave absorption in doped bilayer graphene from linear case. Note, that
in Fig. 6 the doped BG partial cross-sections behavior is the similar to the
intrinsic SG one for\ the certain initial angle $\theta _{0}=\pi /2$ $%
\mathrm{rad}$ only. As is seen from Fig. 6, the envelopes of the partial
cross sections $n\Lambda ^{(n)}$ for $n>1$\ ($n<1$) have many characteristic
maxima (minima) in the doped BG case which is in sharp contrast with the
doped SG case \cite{nanop1} where for the same interaction parameters one
has a single maximum (minimum) in\ absorption (emission) part. Next, we see
from Fig. 6 that with the increase of radiation intensity the absorption
becomes dominant over emission process. Thus, the SB absorption process in
doped BG can be feasible already at moderate radiation intensities. 
\begin{figure}[tbp]
\includegraphics[width=.51\textwidth]{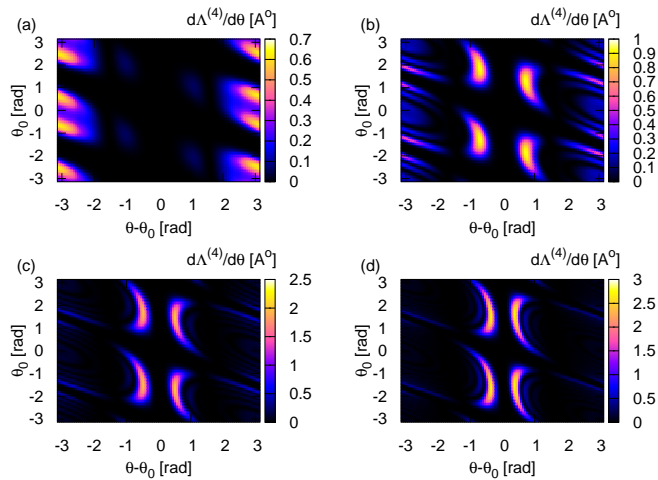}
\caption{{}(Color online) Partial differential cross section $d\Lambda
^{(n)}/d\protect\theta $\ (in angstrom) for photon number $n=4$: ($a$)--($d$%
) correspond to dimensionless field parameters $\protect\chi =0.5,1,1.5$ and 
$2$, respectively.}
\label{55}
\end{figure}
\begin{figure}[tbp]
\includegraphics[width=.51\textwidth]{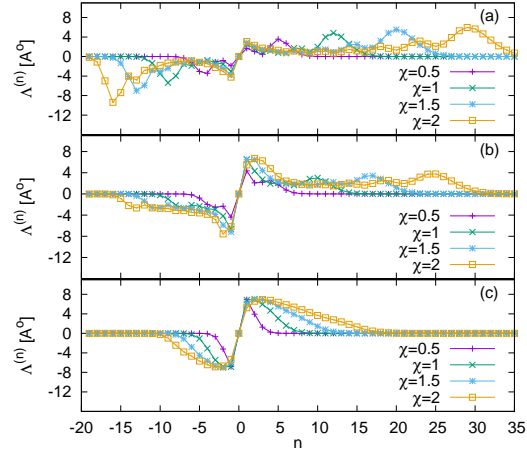}
\caption{{}(Color online) Envelopes of partial absorption-emission cross
sections $n\Lambda ^{(n)}$\ (in angstrom) as a function of the number of
emitted or absorbed photons. The angle $\protect\theta _{0}$ between the
initial electron momentum and wave electric field is taken: (a) $0$, (b) $%
\protect\pi /4$ $\mathrm{rad}$, and (c) $\protect\pi /2$ $\mathrm{rad}$.}
\end{figure}
\begin{figure}[tbp]
\includegraphics[width=.51\textwidth]{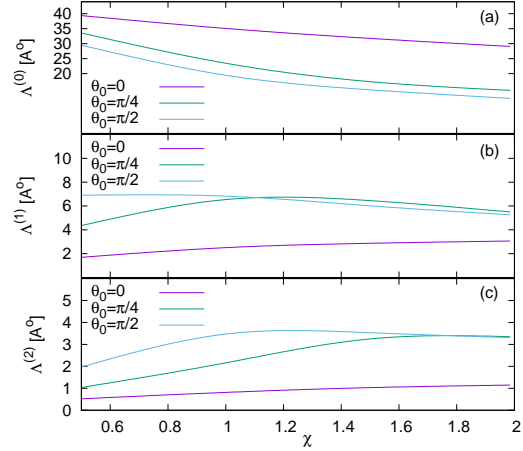}
\caption{(Color online) Partial cross sections $\Lambda ^{(n)}$\ (in
angstrom) versus intensity parameter for several initial angle $\protect%
\theta _{0}$ between the electron momentum and wave electric field. Photon
number is taken: (a) $n=0$ (elastic case), (b) $n=1$, and (c) $n=2$.}
\end{figure}

The total partial cross sections $\Lambda ^{(n)}$ for the elastic, one, and
two photon processes versus the intensity for various initial angles $\theta
_{0}$ are presented in Fig. 7. As is seen from this figure, if the elastic
cross section $\Lambda ^{(0)}$ decreases with the increase of induced
radiation intensity, the partial cross sections $\Lambda ^{(1)}$ and $%
\Lambda ^{(2)}$ grow with the increase of intensity. In the presence of
strong radiation field the elastic cross section is essentially modified.
Hence, the strong coherent radiation field can allow to manipulate with the
electron transport properties of doped BG. Moreover, as is seen from Figs.
1-6 and comparison with the results for SG case \cite{nanop1}, such
procedure can be more effective in BG case.

\section{Conclusion}

The theoretical investigation of the multiphoton SB process in doped BG is
given. Based on the quantum theory, we study the induced scattering of 2D
chiral particles on the charged impurity ions of an arbitrary electrostatic
potential in the Born approximation. To exclude the interband transitions in
doped BG, we have considered high Fermi energies and an external EM wave is
taken to be in the terahertz domain of frequencies. The obtained analytical
formulas for SB in coherent radiation field with linear polarization are
analyzed numerically for a screened Coulomb potential of doped into BG
impurity ions. The latter shows that induced by a power radiation field the
scattering process in BG is strongly nonlinear. In comparison with the SG
case the new behavior has been demonstrated, which may be associated with
the chiral nature of BG and its parabolic energy dispersion. In particular,
it has been obtained many maxima of SB partial absorption cross-section in
doped BG at multiphoton excitation in moderately intense radiation fields.
In addition, the elastic cross section is significantly modified by the pump
wave, which opens new possibilities to manipulate the wave dressed electron
transport properties in doped BG. The comparison with the results of
intrinsic SG case demonstrates that absorption of a strong electromagnetic
radiation in doped BG is of special interest, and this problem will be
addressed in the forthcoming article.

\begin{acknowledgments}
The authors are deeply grateful to prof. H. K. Avetissian for permanent discussions and valuable recommendations.
This work was supported by the RA MES State Committee of Science.
\end{acknowledgments}

\end{document}